\documentclass[11pt,oneside, a4paper]{article}

\ifx\pdfoutput\undefined
\usepackage[dvips,bookmarks=false]{hyperref}	
\else
\usepackage{hyperref}	
\fi
\hypersetup{colorlinks,bookmarksopen,bookmarksnumbered,citecolor=blue,
linkcolor=black,pdfstartview=FitH,urlcolor=blue}

\newcommand{\1}{\mbox{1}\hspace{-0.25em}\mbox{l}}

\usepackage{graphicx}
\usepackage{amssymb}
\usepackage{cite}
\usepackage{bm}
\usepackage{indentfirst}
\usepackage{amsmath}
\usepackage{hhline}
\usepackage{multirow}

\def\be{\begin{equation}}
\def\ee{\end{equation}}
\newcommand{\bea}{\begin{eqnarray}}
\newcommand{\eea}{\end{eqnarray}}

\begin{document}
\begin{titlepage}

\begin{flushright}
LPT-ORSAY-16-38
\end{flushright}

\begin{center}

\vspace{1cm}
{\large\bf 
Electron electric dipole moment in Inverse Seesaw models}
\vspace{1cm}

\renewcommand{\thefootnote}{\fnsymbol{footnote}}
Asmaa Abada$^1$\footnote[1]{asmaa.abada@th.u-psud.fr}
and 
Takashi Toma$^1$\footnote[2]{takashi.toma@th.u-psud.fr}
\vspace{5mm}

{\it
$^{1}$Laboratoire de Physique Th\'eorique, CNRS, \\
Univ. Paris-Sud, Universit\'e Paris-Saclay, 91405 Orsay, France
\vspace*{.2cm} 
}

\vspace{8mm}

\abstract{
We consider the contribution of sterile neutrinos to the electric dipole
 moment of charged leptons in the most minimal realisation of the
 Inverse Seesaw mechanism, in which the Standard Model is extended by
 two right-handed neutrinos and two sterile fermion states. Our study
 shows that the two pairs of (heavy) pseudo-Dirac mass eigenstates can
 give significant contributions to the electron electric
 dipole moment, lying  close to future 
experimental sensitivity if their masses are above the electroweak scale. 
The major contribution comes from  two-loop diagrams with pseudo-Dirac neutrino
 states running in the loops.
 In our analysis we further discuss the possibility of having a
 successful leptogenesis in this framework, compatible with a large
 electron electric dipole moment.}

\end{center}
\end{titlepage}

\renewcommand{\thefootnote}{\arabic{footnote}}
\setcounter{footnote}{0}

\setcounter{page}{2}

\section{Introduction}
The origin of neutrino masses, the relic dark matter abundance  and the
baryon asymmetry of the Universe (BAU) are pressing open questions
calling for extensions of  the Standard Model (SM). One of the minimal
extensions aiming  to provide at least an explanation for the neutrino
oscillation phenomena consists in the addition of right-handed (RH)
neutrinos, singlets under the SM gauge group, giving rise to Dirac mass
terms.   
The fact that the RH neutrinos are sterile fermions implies that they
can  have a Majorana mass  so that  
this simple extension of the SM  corresponds to the embedding of the
seesaw
mechanism~\cite{Minkowski:1977sc,Yanagida:1979as,GellMann:1980vs,Glashow:1979nm,Mohapatra:1979ia, 
Schechter:1980gr,Schechter:1981cv} into the SM. 
The possibility of probing this scenario depends on the mass regime of
the additional sterile states and on the size of their Yukawa couplings to
the active neutrinos. In the ``standard'' type I seesaw, 
 the masses of the RH neutrinos are required to be very large for
 sizeable (natural)  values of the Yukawa couplings, implying that any direct (collider
 observables) or  indirect signals  (low-energy or high-intensity
 observables) are likely impossible to be discovered. 
When the masses of the additional RH states are around or below 
the electroweak scale, these states  can be directly produced  in
colliders and their contribution to low-energy observables can be
important:
this is why  low-scale seesaw
models~\cite{Mohapatra:1986bd,GonzalezGarcia:1988rw,Deppisch:2004fa,Asaka:2005an,Gavela:2009cd,Ibarra:2010xw,Abada:2014vea}
prove to be appealing. Among them, the Inverse
Seesaw mechanism~\cite{Mohapatra:1986bd}, 
the $\nu$MSM~\cite{Asaka:2005an}, the low-scale 
type-I seesaw~\cite{Gavela:2009cd,Ibarra:2010xw} and 
the Linear Seesaw~\cite{Barr:2003nn,Malinsky:2005bi} are examples of models with a rich phenomenology.

Some of the latter  scenarios may also provide a possible explanation to
the dark matter  relic density 
considering, putting forward a  keV-scale sterile neutrino as a viable
candidate~\cite{Asaka:2005an,Abada:2014zra}, 
and/or to the BAU through leptogenesis (via neutrino
oscillations)~\cite{Akhmedov:1998qx,Canetti:2012vf,Canetti:2012kh,Abada:2015rta,Canetti:2014dka, Hernandez:2015wna}.

Other than  the above motivation,  
some of the current neutrino oscillation experiments (reactor~\cite{Mueller:2011nm, Huber:2011wv, Mention:2011rk},
accelerator~\cite{Aguilar:2001ty, AguilarArevalo:2007it, AguilarArevalo:2010wv, Aguilar-Arevalo:2013pmq}  and
Gallium~\cite{Acero:2007su, Giunti:2010zu}) suggest  the 
existence of sterile fermions  with masses in the eV range.
This would  imply that instead of  the three-neutrino mixing scheme (in oscillation phenomena), one
would have a $3+1$-neutrino (or $3 +$more) mixing schemes (see, for
instance,~\cite{Kopp:2013vaa}).

Extensions of the SM with sterile fermions, which accommodate 
oscillation data, may also have an impact on other observables,   
 such as  charged lepton flavour violation (cLFV) in Higgs~\cite{Arganda:2014dta,Deppisch:2015qwa,Banerjee:2015gca}, neutral
 $Z$ boson~\cite{Illana:2000ic,Abada:2014cca, Abada:2015zea}  and meson decays~\cite{Shrock:1980vy,Shrock:1980ct,Atre:2009rg,Abada:2012mc,Abada:2013aba}. Sterile fermions may also contribute to
 lepton flavour conserving observables such as the dipole moments of
 charged leptons~\cite{deGouvea:2005jj,Abada:2014nwa,Abada:2015trh}. Coupling to the active neutrinos, the sterile fermions
 may also add new sources of CP violation to the already existing one of
 the SM, and can  thus provide new contributions to different CP-odd observables, among them  electric dipole moments (EDMs)~\cite{deGouvea:2005jj}.

In a  recent work~\cite{Abada:2015trh}, we studied the impact of sterile fermions on the electric dipole
moments  of charged
 leptons in the context of the SM extended by an arbitrary number\footnote{Being SM gauge singlets, there is no constraint  on their (generation)
number from anomaly cancellation.} of  sterile
 neutrinos - without necessarily invoking a mechanism of neutrino mass generation - and  we have shown that in order to have a non-vanishing
 contribution, the minimal number of sterile fermion states that must be added to the SM  field content is two.  
 In particular, in the framework of this ad-hoc construction, the latter states 
can  give significant contributions to the charged lepton EDMs, some of them lying within future 
experimental sensitivity if their (non-degenerate) masses
are both above the electroweak scale. 
The Majorana nature of the neutrino states is also an  important ingredient in
order to allow
 for significative contributions to the charged lepton EDMs.  
 
In this work, we consider the electron EDM in a specific seesaw realisation, the
Inverse Seesaw (ISS) in its minimal version,  which offers the
possibility of accommodating  the smallness of the light (mostly active) neutrino masses
for a comparatively low seesaw scale, but still with large values of
the Yukawa couplings. 

In~\cite{Abada:2014vea},  
it was shown that it is possible to construct several minimal 
ISS realisations  that can 
reproduce the correct neutrino mass spectrum while fulfilling all
phenomenological constraints, each exhibiting distinct features.  This allowed to identify  a truly minimal
ISS realisation denoted ``(2,2) ISS'' model, where the SM is extended
by 
two RH neutrinos and two sterile states. This configuration leads to a
3-flavour mixing scheme in the normal hierarchy for the light  neutrinos, 
and  two pairs of (heavy) pseudo-Dirac mass eigenstates. 

Compared to the ad-hoc construction previously mentioned, in the (2,2) ISS realisation, the additional CP phases and the peculiar heavy spectrum (two pairs of pseudo-Dirac states) may lead to different prospects concerning the charged lepton EDMs. 

The present study shows that the  pseudo-Dirac states can give significant contributions to the electron EDM,   close to  the future 
experimental sensitivity, should the pseudo-Dirac masses be above the electroweak scale.  
We have shown that, contrary to the ad-hoc model, the two-loops diagrams relying on the Majorana nature of the exchanged neutrinos turn out to be suppressed.
Interestingly, due to the structure of the spectrum and of the lepton mixings, in the (2,2) ISS,  the major contribution to the EDMs  arises  from the diagrams with Dirac-like fermions  in the loops. 
Furthermore, we have estimated the maximal enhancement factor to the charged lepton EDMs in the case of a generic $(N,N)$ ISS realisation, with  $N>2$.
 In this work we also discuss the possibility of having a successful (thermal) 
 leptogenesis in this ISS framework in the regimes associated with significant contributions to the electron EDM.

The paper is organised as follows:  in Section~\ref{Sec:Model}, after describing 
the (2,2) ISS realisation,  we discuss in detail 
 the degrees of freedom associated to  the CP-violating phases, the mass regimes of the sterile states and the active-sterile mixing angles. 
 We also summarise the relevant constraints on these  extensions of the SM. 
Section~\ref{Sec:EDM} is devoted to the charged lepton EDMs,  including a thorough  discussion of the
several two-loop diagrams. 
The impact of  sterile neutrino contributions to the EDMs within this minimal ISS realisation is numerically evaluated and presented in  Section~\ref{Sec:Num}, while the analytical determination of the EDM is summarized in the Appendix. 
Our final remarks  and  discussion are collected  in Section~\ref{Sec:
Conclusions}. We summarize our results in Section~\ref{Sec: Summary}.

\section{The (2,2) ISS model}
\label{Sec:Model}

\subsection{The neutrino mass matrix}
As argued in~\cite{Abada:2014vea}, the minimal realisation of the Inverse Seesaw model requires the
addition of  two right-handed neutrinos $N_i$,  and
two singlet fermions $s_i$ to the SM field content. Assigning the same  lepton numbers ($L=+1$)  to $N_i$ and $s_i$ 
allows for a small $\Delta L=2$ lepton number violating (LNV) mass
parameter $\mu$ and $m$, corresponding to  Majorana masses in the sterile
sector. This leads to the following neutrino mass terms in  the Lagrangian 
\bea
- \mathcal{L}_{m_\nu} = n_L^T\ C\ {M}\ n_L + \mathrm{h.c.},
\eea
where
\bea
\begin{array}{ccc}
n_L \equiv  \left(  \nu_L^1,\nu_L^2, \nu_L^3, N_1^{c},N_2^{c} , s_1 ,s_2
	    \right)^T, & \rm{and} &  C = i \gamma^2\gamma^0\ . 
\end{array}
\eea
The mass matrix $\mathcal{M}$ can be written  as:
\be\label{m22full}
{M} = \left( \begin{array}{ccccccc} 
0&0&0&d_{11}&d_{12}&0&0\\
0&0&0&d_{21}&d_{22}&0&0\\
0&0&0&d_{31}&d_{32}&0&0\\
d_{11}&d_{21}&d_{31}&m_{11}&m_{12}&n_{11}&n_{12}\\
d_{12}&d_{22}&d_{32}&m_{12}&m_{22}&n_{21}&n_{22}\\
0&0&0&n_{11}&n_{21}&\mu_{11}&\mu_{12}\\
0&0&0&n_{12}&n_{22}&\mu_{12}&\mu_{22}
 \end{array} \right)\ . 
\ee

One can always choose a basis in which the unphysical parameters are
reabsorbed via appropriate redefinitions of the fields; one of the  possible choices of  basis - which we have adopted \cite{Abada:2014vea} - is summarised in Table \ref{basis-22}, leading to
24 physical parameters. 
\begin{table}[htbp]
\begin{center}
\begin{tabular}{|c|c|c|c|}
\hline
Matrix & \# of moduli & \# of phases &  Total \\
\hline
Diagonal and real ${\mathfrak{m_\ell}}$ & $3$ & $0$ & $3$ \\
$d$ with one real column & $6$ & $3$ & $9$ \\
 $m$ & $3$ & $3$ & 6 \\
Real and diagonal $n$ & $2$ & $0$ & 2 \\
$\mu$ with real diagonal & $3$ & $1$ & 4\\
\hline
Total & $17$ & $7$ & $24$\\
\hline 
\end{tabular}
\end{center}
\caption{Example of a basis in which the number of  parameters matches
 the number of physical degrees of freedom (${\mathfrak{m_\ell}}$ corresponds to
 the charged lepton masses).}
\label{basis-22}
\end{table}

In the following we will use this theoretical framework , denoted ``(2,2) ISS''  model. Moreover, we will neglect the  mass parameters $m_{ij}$, which induce subdominant effects when compared to the  entries $\mu_{ij}$ in the mass matrix of 
Eq.~(\ref{m22full}). 
Indeed, both these lepton number violating  mass matrices  can be dynamically generated  as done  in
the general original formulation of the Inverse Seesaw mechanism~\cite{Mohapatra:1986bd},
in which the smallness of the  $\mu$ matrix elements  was attributed to 
supersymmetry breaking effects in a (superstring inspired) $E_6$
scenario. For instance, in the context of a non-supersymmetric $SO(10)$
model, which contains the remnants of a larger $E_6$ group, the mass matrix $ \mu$ is
generated at two-loop level while the matrix $m$ is generated at higher order, thus justifying
the smallness of its entries compared to the ones of $\mu$~\cite{Ma:2009gu}.
Once  the lepton number violating terms $m_{ij}$ are neglected, the mass
matrix is  reduced to 
\begin{equation}
\label{iss22simple}
M=\left(
\begin{array}{ccccccc}
0 & 0 & 0 & d_{11} & d_{12} & 0 & 0\\
0 & 0 & 0 & d_{21} & d_{22} & 0 & 0\\
0 & 0 & 0 & d_{31} & d_{32} & 0 & 0\\
d_{11} & d_{21} & d_{31} & 0 & 0 & n_1 & 0\\
d_{12} & d_{22} & d_{32} & 0 & 0 & 0 & n_2\\
0 & 0 & 0 & n_1 & 0 & \mu_{11} & \mu_{12}\\
0 & 0 & 0 & 0 & n_2 & \mu_{21} & \mu_{22}\\
\end{array}
\right)\ ,
\end{equation}
where the sub-matrix $d$ is parametrised by $6$ moduli and $3$ CP phases, $n$ via 
$2$ moduli, while  $\mu$ includes $3$ moduli and $1$ CP phase. 
Thus the mass matrix $M$ totally is defined by $11$ moduli and $4$ CP
phases. 
Since the determinant of the mass matrix in Eq.~(\ref{iss22simple})
vanishes, the lightest mass eigenvalue is zero in the minimal Inverse Seesaw
model.  The diagonalisation of the mass matrix in Eq.~(\ref{iss22simple}) leads at leading order to
three light (almost active) neutrinos (systematically in the  normal hierarchy
ordering, as found in  \cite{Abada:2014vea}), and  to two
pseudo-Dirac pairs containing the mostly sterile eigenstates,  with mass
differences of the order of the LNV entries of the $\mu$
sub-matrix.\footnote{In the limit in which  lepton number is conserved
(i.e. $\mu\to 0$), these states become Dirac particles.}

The weak charged current Lagrangian for the leptons 
is modified as 
\bea
- \mathcal{L}_\text{cc} = \frac{g}{\sqrt{2}} U_{\alpha i} 
\overline{\ell_\alpha} \gamma^\mu P_L \nu_i  W_\mu^- + \, \text{h.c.}\,,
\eea
where $U_{\alpha i}$ is the unitary lepton mixing matrix, 
$i = 1, \dots, 7$ denotes the physical neutrino states
and $\alpha = e, \mu, \tau$ the flavour of the charged leptons. 
In the case of three neutrino generations,  $U$ would correspond 
to the ($3\times 3$ unitary) PMNS matrix, $U_\text{PMNS}$. 
The mixing between the left-handed leptons, here denoted by $\tilde
U_\text{PMNS}$, 
now corresponds to a $3 \times 3$ block of the $7\times 7$ unitary matrix $U$, which can be parametrised as
\begin{equation}\label{eq:etatilde}
U_\text{PMNS} \, \to \, \tilde U_\text{PMNS} \, = \,(\1 - \eta)\, 
U_\text{PMNS}\,,
\end{equation}
where the  matrix $ \eta$ encodes the deviation of $\tilde
U_\text{PMNS}$ from unitarity~\cite{Schechter:1980gr,Gronau:1984ct}.
It also convenient to introduce the invariant quantity $\tilde\eta= 1-|\mathrm{Det}( \tilde U_\text{PMNS})|$, particularly useful to illustrate the effect of  the active-sterile mixings.

\subsection{Constraints}\label{sec:constraint}

Depending on their masses and on the mixings with the active (light)
neutrinos, the sterile states are severely constrained from several
observations. Firstly, these extensions should account for oscillation
data. In our analysis we have required compatibility with the best fit
intervals for a normal hierarchical light
spectrum~\cite{Gonzalez-Garcia:2014bfa},  
\begin{eqnarray}
0.270\leq\sin^2\theta_{12}\leq0.344\,,\qquad
0.382\leq\sin^2\theta_{23}\leq0.643\,,\qquad
0.0186\leq\sin^2\theta_{13}\leq0.0250\,,\\
7.02\leq\frac{\Delta{m}_{21}^2}{10^{-5}~\mathrm{eV}^2}\leq8.09\,,\qquad
2.317\leq\frac{\Delta{m}_{31}^2}{10^{-3}~\mathrm{eV}^2}\leq2.607\,.
\hspace{2cm}
\end{eqnarray}
(As discussed in~\cite{Abada:2014vea}, for such a minimal realisation of the Inverse
      Seesaw model, 
 an inverted hierarchy is strongly disfavoured.) We also notice that 
the non-unitarity of the $\tilde U_{\text{PMNS}}$ (sub-matrix) 
is constrained by a number of observations, as discussed
in~\cite{Antusch:2008tz, Antusch:2014woa}. 

Further constraints on the active-sterile mixings and on the sterile
neutrino masses can
be inferred from current bounds arising from neutrinoless double beta
decays~\cite{Benes:2005hn}. In the present scenario, the relevant effective neutrino mass is given by~\cite{Blennow:2010th,Abada:2014nwa}
\begin{equation}
m_{ee}=\sum_{i=1}^{7}U_{ei}^2\frac{p^2m_i}{p^2-m_i^2}\,,
\end{equation}
where $p^2=-\left(125~\mathrm{MeV}\right)^2$. 

Several experiments (like GERDA~\cite{Agostini:2013mzu}, 
EXO-200~\cite{Auger:2012ar,Albert:2014awa}, and 
KamLAND-ZEN~\cite{Gando:2012zm}) have put constraints on $|m_{ee}|$, which translate into bounds on combinations of $U_{ei}^2 m_i$,
$i=4,.., 7$. In our numerical study,  we have imposed that our solutions
always comply with the (conservative) experimental bound $|m_{ee}|
\lesssim 0.01$ eV.

Particularly relevant for the case of a large sterile mass regime is
the perturbative unitarity bound: 
if the additional sterile fermions are sufficiently heavy  to 
decay into a $W$ boson and a charged lepton, or into an active neutrino
and either a $Z$ or a Higgs boson, their decay widths should comply with
the perturbative unitarity condition~\cite{Chanowitz:1978mv,Durand:1989zs,Korner:1992an,
Bernabeu:1993up,Fajfer:1998px,Ilakovac:1999md,Abada:2014cca}. 
In this case, and since the dominant decay mode of the (mostly) sterile neutrinos,
$\nu_{4...7}$, would be $\nu_i\to\ell_\alpha^\mp W^\pm$, 
 their decay width  should  comply with the perturbative
 unitary bound\footnote{Another common criterion of perturbativity is
 that the couplings  should be less than $\sqrt{4\pi}$. This
 criterion also gives a bound similar to  Eq.~(\ref{eq:pert}).}:
\bea 
\frac{\Gamma_{\nu_i}}{m_{i}}\, < \, \frac{1}{2}\quad \text{where}
\quad \Gamma_{\nu_i}\approx 
\frac{g_2^2 m_i^3}{16\pi m_W^2}\sum_{\alpha}\left|U_{\alpha
i}\right|^2\ , \, (i =4,..,7)\, , \
\eea
which translates into an upper bound on the sterile neutrino masses as
follows,  
\begin{equation}
m_i\lesssim873~\mathrm{GeV}
\left(\sum_{\alpha}|U_{\alpha i}|^2\right)^{-1/2}\,.
\label{eq:pert}
\end{equation}

Important bounds arise from electroweak precision tests; the
active-sterile mixings are constrained from 
observables such as the $W$ boson decay width,
the $Z$ invisible decay, meson decays and the non-unitarity of the
$3\times3$ sub-matrix  ($\tilde U_{\text{PMNS}}$) of $U_{ij}$.
For $m_i<m_{W},m_Z$ $(i=4-7)$, the most important constraints arise
from the $W$ decay and the $Z$ invisible decay; for the 
neutrino mass range $3~\mathrm{GeV}\lesssim m_i\lesssim
90~\mathrm{GeV}$, the strongest constraints are those of 
the DELPHI~\cite{Abreu:1996pa} and
 L3~\cite{Adriani:1992pq} Collaboration. 

Additional sterile states might also lead to the violation of 
lepton flavour universality, as arising from  
meson decays such as $\pi^+\to \ell_\alpha^+\nu_\alpha$ and
$K^+\to \ell_\alpha^+\nu_\alpha$~\cite{Abada:2013aba,
Abada:2012mc,Asaka:2014kia}. 
Lepton flavour violating processes also provide important constraints
on the sterile fermion parameter space; in particular, the bounds from
the muon-electron sector lead to the strongest constraints, which are 
\begin{equation}
\mathrm{Br}\left(\mu\to{e}\gamma\right)\leq5.7\times10^{-13},\quad
\mathrm{Br}\left(\mu\to{e}\overline{e}e\right)\leq1.0\times10^{-12},\quad
\mathrm{Cr}\left(\mu-e, \mathrm{Au}\right)\leq7\times10^{-13}\,,
\end{equation}
as obtained, respectively, by~\cite{Adam:2013mnn},~\cite{Bellgardt:1987du} and \cite{Bertl:2006up}. 

Direct searches at LEP have put strong constraints on  
sterile neutrinos whose masses are
$m_i\lesssim\mathcal{O}(100)~\mathrm{GeV}$.
The relevant process is
$e^+e^-\to\nu_i\nu_j^*\to\nu_ie^\pm W^\mp$ where $i\leq3$ and 
$j\geq4$, which violates lepton number conservation. 
This has allowed to exclude 
certain regimes of the mixing angles $|U_{\alpha i}|$~\cite{Deppisch:2015qwa}. 
 Searches at the LHC for a same sign
di-lepton 
channel $pp\to {W^\pm}^*\to\ell^{\pm}\nu_i\to\ell^\pm\ell^\pm jj$ (where
$i\geq4$ and $j$ denotes a jet), have led to further bounds for 
$m_i\gtrsim\mathcal{O}(100~\mathrm{GeV})$: for values of the 
integrated luminosity of $20~\mathrm{fb}^{-1}$ at  
$\sqrt{s}=8~\mathrm{TeV}$, LHC data already
allows to constrain the mixing angle $|U_{\alpha i}|$
for  sterile neutrino masses up to
$500~\mathrm{GeV}$~\cite{Dev:2013wba,Das:2014jxa}.

\section{Electric Dipole Moments}\label{Sec:EDM}

\begin{figure}[t]
\begin{center}
\includegraphics[scale=0.8]{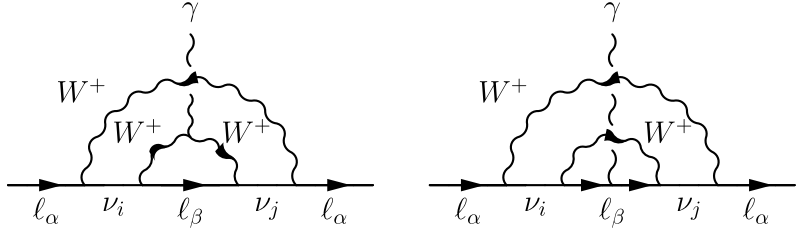}
\caption{Diagrams for charged lepton EDMs.}
\label{fig:edm}
\end{center}
\end{figure}

In Inverse Seesaw models, the charged lepton  EDMs are induced at two-loop level 
as shown in Refs.~\cite{deGouvea:2005jj,Abada:2015trh}. 
In general, there is a very large number ($\sim 100$) of diagrams contributing to the charged lepton 
EDM\footnote{
The number of  relevant diagrams to EDMs depends on the chosen gauge. 
If a non-linear gauge is taken, the number of diagrams is 
considerably reduced~\cite{Gavela:1981ri, Belanger:2003sd}.}; however and noticing that the heavy (sterile) neutrinos providing the dominant contributions form pseudo-Dirac pairs, the number 
of  diagrams that must be evaluated  can be significantly reduced, as we will presently discuss.
 In a previous work~\cite{Abada:2015trh}, we have discussed the several possible contributing diagrams in the case of an ad-hoc model corresponding to the Standard Model extended by N sterile states, considering both Majorana and Dirac fermion contributions. 
There was, in the Feynman gauge,  in total 44 diagrams of type of Fig.~\ref{fig:edm}, as well as  96 additional diagrams corresponding to  $Z$ and to the Higgs bosons mediation. 
In this previous study we have computed the diagrams and discussed the relevance of each contribution, distinguishing also the Dirac contribution from the Majorana one. It is worth noticing that in the Inverse seesaw framework, the number of diagrams that must be evaluated  is significantly reduced, as we will presently discuss.

In ``effective'' neutrino models (corresponding to minimalistic ad-hoc constructions where the SM is extended by $n$ sterile fermions),  
the EDM of the electron can be 
formally written as~\cite{Abada:2015trh}
\begin{equation}
d_e=-\frac{g_2^4\,e\,m_e}{4(4\pi)^2m_W^2}\sum_{\beta}\sum_{i,j}
\Bigl[
J_{ije\beta}^MI_M(x_i,x_j)
+J_{ije\beta}^DI_D(x_i,x_j)
\Bigr],
\label{eq:edm}
\end{equation}
where $x_{i,j}\equiv m_{i,j}^2/m_W^2$ and
$m_{\alpha,\beta}^2/m_W^2\ll1~(\alpha,\beta=e,\mu,\tau)$, 
$I_M$ and $I_D$ are the loop functions whose analytic expressions (for the
dominant Majorana contributions) can be found in~\cite{Abada:2015trh}. 
The (CP-odd)  factors $J_{ij\alpha\beta}^M$ and $J_{ij\alpha\beta}^D$
are defined by
\begin{equation}
J_{ij\alpha\beta}^M\equiv\mathrm{Im}\left(U_{\alpha{j}}U_{\beta{j}}U_{\beta{i}}^*U_{\alpha{i}}^*\right),\qquad
J_{ij\alpha\beta}^D\equiv\mathrm{Im}\left(U_{\alpha{j}}U_{\beta{j}}^*U_{\beta{i}}U_{\alpha{i}}^*\right).
\label{eq:phase}
\end{equation}
Similar expressions hold for the EDMs of the other charged leptons ($\mu$ and $\tau$); 
however we do not consider them here since the predicted EDMs for $\mu$ and
$\tau$ are extremely
tiny compared to the sensitivities of future experiments. 
The first term in the brackets in Eq.~(\ref{eq:edm}) represents a 
contribution which reflects the Majorana nature of the sterile fermions, while
the second term corresponds to a generic Dirac fermion contribution.  
As one can see from the definition of the factors
$J_{ij\alpha\beta}^M$ and $J_{ij\alpha\beta}^D$, these  are totally
anti-symmetric in terms of $i\leftrightarrow j$, implying that a non-vanishing EDM requires contributions from two neutrino states with $i\neq j$. 
 The loop functions $I_M(x_i,x_j)$ and
$I_D(x_i,x_j)$ should also be anti-symmetric under $i\leftrightarrow j$.
As a result, one can see that the EDM itself is fully symmetric under
$i\leftrightarrow j$.

 Such a formulation is valid for any extension of the SM involving sterile fermions, as is the case of  type-I,  Inverse and Linear Seesaw models.  
In Inverse Seesaw models (with equal number of $N_i$ and $s_i$), the heavy sterile neutrinos form pseudo-Dirac
fermion pairs. Moreover, within a pseudo-Dirac pair, the states are highly degenerate in mass (recall that their non-degeneracy is proportional to the entries of the $\mu$ matrix).  
 The  Majorana contribution in Eq.~(\ref{eq:edm}) is thus very suppressed when 
compared to the Dirac one.\footnote{ 
We have numerically confirmed that the phase factor for the Majorana
contribution $J^M_{ij\alpha\beta}$ is indeed highly 
suppressed as shown in Fig.~\ref{fig:j_fac} in the next section.} 
Taking into account this fact, the expression of the electron EDM for the case of 
Inverse Seesaw models is simplified to 
\begin{equation}
d_e\approx-\frac{g_2^4\,e\,m_e}{4(4\pi)^2m_W^2}\sum_{\beta}\sum_{i,j}
J_{ije\beta}^DI_D(x_i,x_j). 
\label{eq:edm1}
\end{equation}
In addition, in the case of the ``(2,2)''  minimal Inverse Seesaw model, the two
pairs of heavy sterile neutrinos, $(\nu_4,\nu_5)$ and $(\nu_6,\nu_7)$,  are
nearly degenerate respectively,  with  $m_{4,5}<m_{6,7}$. 
Taking into account this fact and unitarity of the mixing matrix ($UU^\dag=\1$), 
the formula of the electron EDM can be further simplified by
\begin{equation}
d_e\approx-\frac{g_2^4\,e\,m_e}{2(4\pi)^2m_W^2}J^DI_D^\prime(x_4,x_6), 
\label{eq:edm2}
\end{equation}
where the loop function $I_D^\prime$ and phase factor $J^D$ are defined by
\begin{eqnarray}
I_D^\prime(x_4,x_6)
\hspace{-0.2cm}&\equiv&\hspace{-0.2cm}
I_D(0,x_4)
-I_D(0,x_6)
+I_D(x_4,x_6),\label{IprimeD}\\
J^D
\hspace{-0.2cm}&\equiv&\hspace{-0.2cm}
\sum_{\beta}
\Bigl[J_{46e\beta}^D+J_{47e\beta}^D+J_{56e\beta}^D+J_{57e\beta}^D\Bigr]\ ;
\end{eqnarray}
and the factor $2$ difference between Eq.~(\ref{eq:edm1}) and
~(\ref{eq:edm2}) arises from having the EDM expression totally symmetric
under $i\leftrightarrow j$.

The electron EDM has been experimentally searched for by ACME
     Collaboration~\cite{Baron:2013eja}, and the current upper bound is given by
\begin{equation}
|d_e|/e\leq8.7\times10^{-29}~\mathrm{cm}.
\end{equation}
 The comparison of the above bounds with  Eq.~(\ref{eq:edm2}) allows to set the limit $|J^DI_D^\prime(x_4,x_6)|\lesssim9.7\times10^{-5}$. 
The upper bound is expected to be improved to 
$|d_e|/e\lesssim10^{-30}~\mathrm{cm}$ by the upgraded ACME
Collaboration~\cite{acme:next_generation}.

\section{Numerical results}\label{Sec:Num}

\begin{figure}[t]
\begin{center}
\includegraphics[scale=0.6]{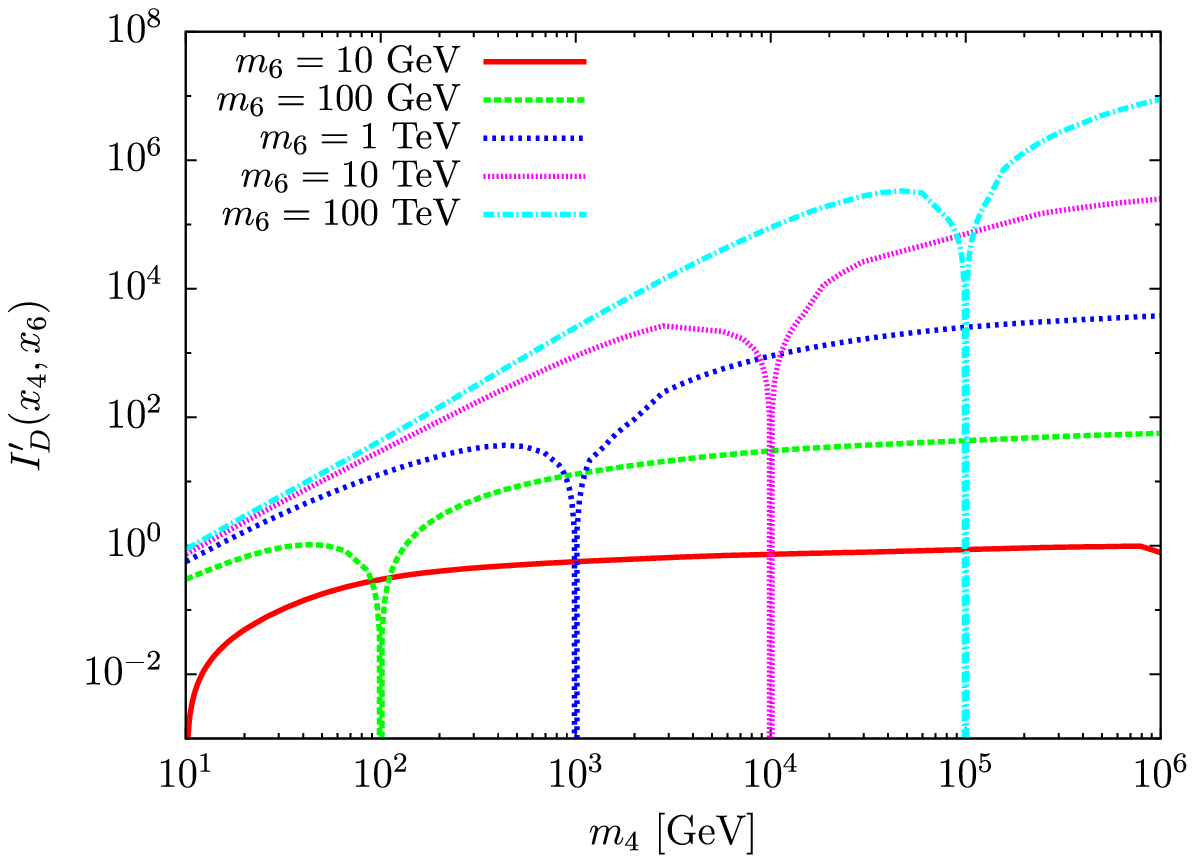}
\hspace{0.3cm}
\includegraphics[scale=0.6]{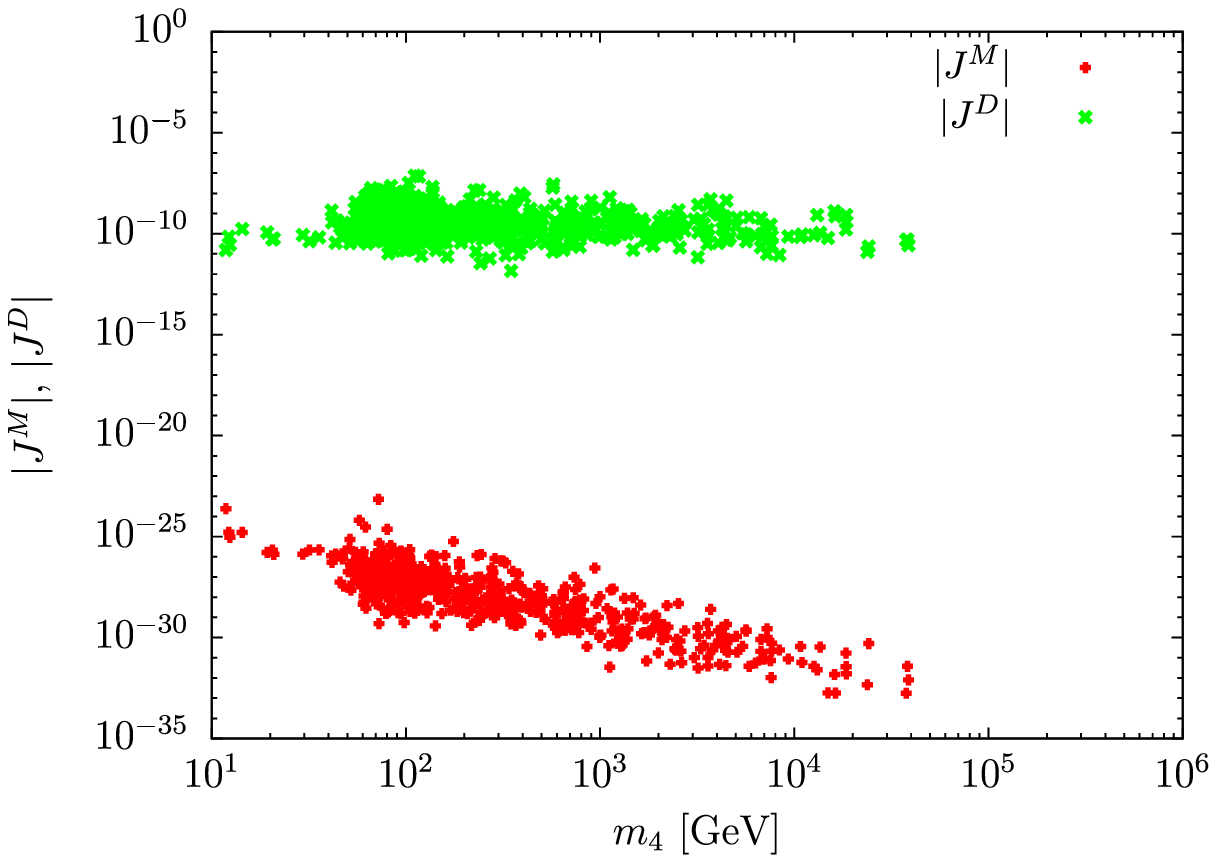}
\caption{On the left, loop function $|I_D^\prime(x_4,x_6)|$ for several fixed values of $m_6$; on the right,  CP-odd  factors
 $|J^M|$ and $|J^D|$ corresponding to points in parameter space satisfying all the 
 constraints discussed in Section~\ref{Sec:Model}.}
\label{fig:j_fac}
\end{center}
\end{figure}

The most difficult  part of the computation of the charged lepton EDMs is the evaluation of the
loop function $I_D(x_i,x_j)$ in Eq.~(\ref{eq:edm1}). This can be done with FeynCalc~\cite{Mertig:1990an}
and the analytical expressions that we we have obtained are given in the  Appendix, where the derived
analytical formulas are written by multiple integrals.
The loop function $I_D^\prime$ of Eq.~(\ref{IprimeD}) has been numerically evaluated, and some illustrative examples have been collected in the left panel of Fig.~\ref{fig:j_fac}, in which we display $I_D^\prime$ as a function of $m_4$ (nearly degenerate with $m_5$, $m_5\approx m_4$) for several fixed values of $m_6$ ($m_7\approx m_6$). Recall that this degeneracy is a consequence of the pseudo-Dirac nature of the heavy spectrum. 
 The sign of the loop function changes at $m_4=m_6$, corresponding to the ``singularities'' in the absolute value of $I_D^\prime$, as can be seen on the figure, and the loop function becomes approximately flat for regimes where 
$m_4\gg m_6$, i.e., for a strongly hierarchical heavy spectrum. 
Although the loop function could  be larger for heavier sterile
neutrinos, these regimes  are theoretically and experimentally
constrained,  in particular by the perturbative unitarity bound and 
by constraints arising from cLFV observables. 

A full analysis  requires to carry a numerical diagonalisation of the $7\times 7 $ neutrino mass matrix of Eq.~(\ref{iss22simple}). In order to account for neutrino oscillation data, the Inverse Seesaw mechanism parameters must fulfill the following condition 
\begin{equation}\label{seesaw-condition}
|\mu|\ll|d|\ll|n|,
\end{equation}
where $\mu$, $d$ and $n$ are the elements of the sub-matrices in
Eq.~(\ref{iss22simple}). 
In addition, since the light neutrino mass matrix can be approximately given by  
\begin{equation}
m_\nu^\mathrm{light}\simeq  
d\ (n^{-1})^T\ {\mu}\ n^{-1} \ d^T\,,
\end{equation}
the lepton number violating parameter $\mu$ would be given
by $\mu\sim m_\nu^\mathrm{light}n^2/d^2$. 
The ratio $d/n$ is related with the non-unitarity of the $\tilde
U_\text{PMNS}$ matrix
and is thus experimentally constrained. 
To satisfy the latter constraints, the ratio should obey 
$d/n\lesssim0.1$~\cite{Abada:2014vea}. 
Accordingly, we take the following intervals of the different entries of the neutrino mass matrix, 
\begin{eqnarray}
1~\mathrm{GeV}\leq n_{i}\leq10^{7}~\mathrm{GeV},\quad
10^{-3}\leq
\frac{\left|d_{ij}\right|}{\mathrm{max}[n_i]}\leq10^{-1},
\hspace{2.5cm}\\
\frac{\mathrm{min}[n_i^2]}{\mathrm{max}[|d_{ij}|^2]}\leq
\frac{\mu_{11}}{10^{-10}~\mathrm{GeV}}\leq\frac{\mathrm{max}[n_i^2]}{\mathrm{min}[|d|_{ij}^2]},\quad
10^{-6}\leq\left|\frac{\mu_{12}}{\mu_{11}}\right|\leq10^{-3},\quad
10^{-1}\leq\left|\frac{\mu_{22}}{\mu_{11}}\right|\leq10. 
\end{eqnarray}

All the parameters in the mass matrix are randomly taken in the above ranges, 
and the different CP  phases are  also randomly varied in the $[ 0, 2\pi]$ interval.

We display on the right panel of  Fig.~\ref{fig:j_fac}  the phase factors $|J^D|$
and $|J^M|$ computed  using the data points complying with all the
constraints discussed in Section~\ref{sec:constraint}. 
One can observe from the figure that the Dirac contribution, $|J^D|$, is
dominant as  expected from our previous discussion; this justifies having the dominant contribution arising from the diagrams displayed in Fig.~\ref{fig:edm} (corresponding to the contribution of  generic Dirac neutrinos in the loops).

The maximum value of the  factor $|J^D|$ is approximately given by
\begin{equation}
|J^D_\mathrm{max}|\sim10^{-5}\times\left(\frac{\mathrm{GeV}}{m_4}\right).
\end{equation}

\begin{figure}[t]
\begin{center}
\includegraphics[scale=0.6]{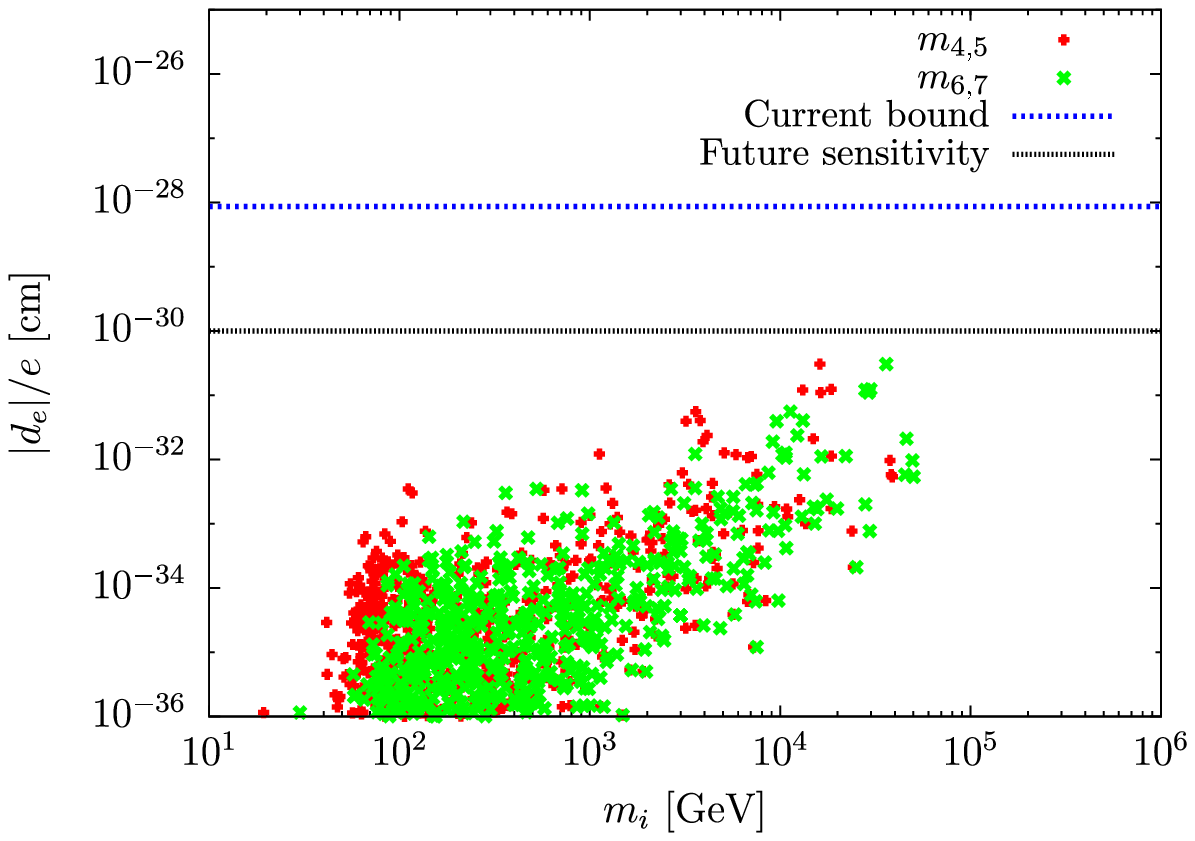}
\hspace{0.5cm}
\includegraphics[scale=0.6]{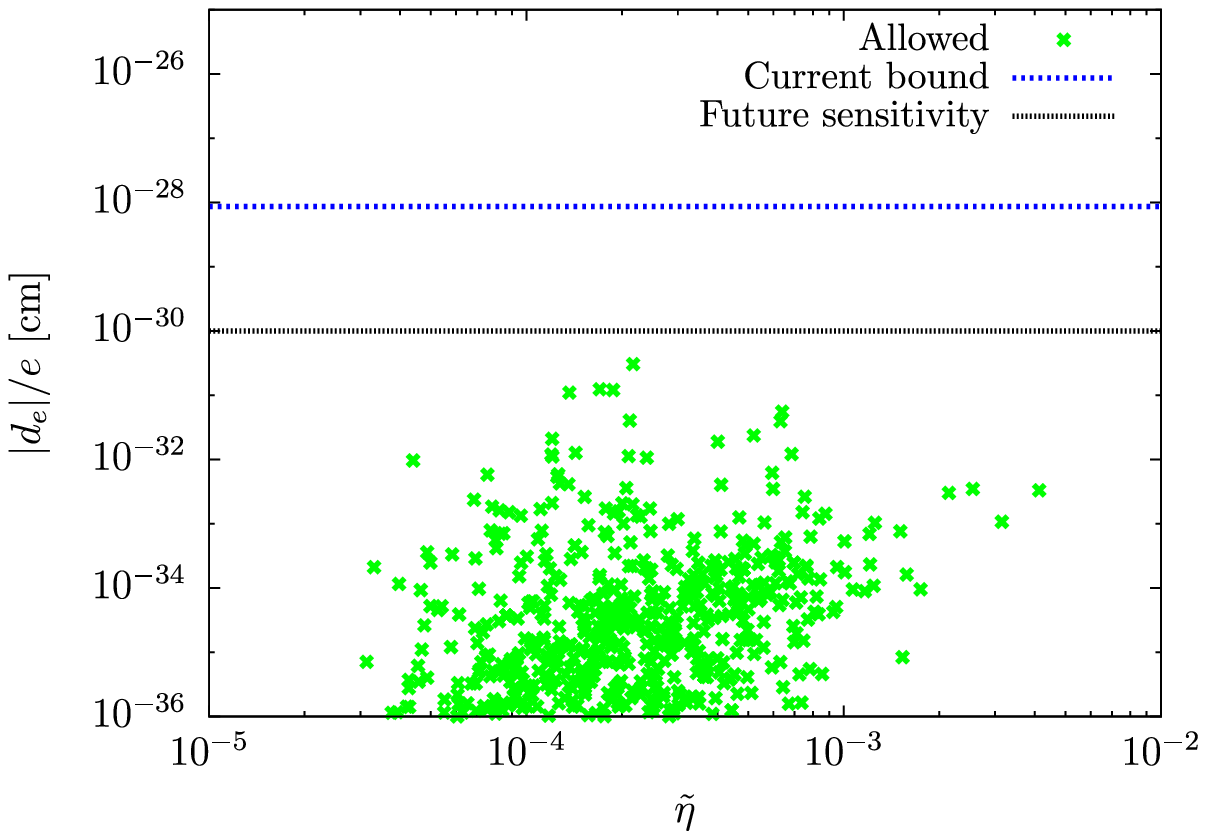}
\caption{Electron EDM as a function of $m_i$ (left) and $\tilde{\eta}$
 (right), all the points displayed comply with the   constraints discussed in Section~\ref{sec:constraint}.}
\label{fig:result1} 
\end{center}
\end{figure}

Combining the above quantities, one can compute the electron EDM, and
the results are shown in Fig.~\ref{fig:result1} as a function of $m_i$ (left) and 
$\tilde{\eta}$ (right), where $\tilde\eta= 1-|\mathrm{Det}( \tilde U_\text{PMNS})|$. As can be seen,  the maximum value of the predicted electron EDM is
$|d_e^\mathrm{\:max}|/e\sim5\times10^{-31}~\mathrm{cm}$, lying 
two orders of magnitude below the current experimental bound, 
$|d_e|/e\leq8.7\times10^{-29}~\mathrm{cm}$,  and thus marginally short of 
the future sensitivity, $|d_e|/e\sim10^{-30}~\mathrm{cm}$.

Contrary to the  previous study~\cite{Abada:2015trh}, where an ad-hoc ``3 + 2 toy'' construction was used (two sterile fermions added to the SM field content), the present framework (the (2,2) ISS model) leads to a much more constrained scenario: firstly, the 
seesaw condition of Eq.~(\ref{seesaw-condition}) strongly constrains the different couplings; secondly, the very nature of the heavy spectrum (pseudo-Dirac pairs) reduces the set of contributing diagrams. Finally, we stress that experimental constraints, as is the case of $\mu\to e\gamma$, are very severe in the mass regime where the most important contributions to the EDMs are expected to arise (above the electroweak scale). We do not dismiss the possibility that the (2,2) ISS model could eventually account for larger values of the electron EDM, but this would require an important amount of fine-tuning between the relevant parameters - and in the present study we chose not to explore such fine-tuned scenarios. 

We have also computed the charged lepton anomalous magnetic moment which
is given by
\begin{equation}
\Delta{a}_\ell=-\frac{4\sqrt{2}G_Fm_\ell^2}{(4\pi)^2}
\sum_{i=4}^{7}|U_{\ell i}|^2G_{\gamma}\left(\frac{m_i^2}{m_W^2}\right)\,, 
\end{equation}
where the loop function $G_{\gamma}(x)$ is given by 
\begin{equation}
G_\gamma(x)=\frac{x-6x^2+3x^3+2x^4-6x^3\log{x}}{4(1-x)^4}\,.
\label{eq:ggamma}
\end{equation}
 For the muon anomalous magnetic moment, the new
contribution is negative and cannot explain the discrepancy of
$3.5\sigma$ deviation between the
SM prediction and the experimental value given by 
$\Delta{a}_\mu=2.88\times10^{-9}$~\cite{Agashe:2014kda}. 
For the electron anomalous magnetic moment, we can na\"\i vely expect   from $\Delta{a}_\mu$ - assuming a common New Physics origin to both discrepancies -  
 that the deviation between the theoretical prediction  of the  SM and the corresponding experimental value will be of the  order of $\Delta{a}_e\sim6.7\times10^{-14}$, 
by scaling the value for the 
muon by $m_e^2/m_\mu^2$. 
However, the current difference between experimental observation  and the
SM prediction is  $\Delta{a}_e=8.2\times10^{-13}$~\cite{Aboubrahim:2014hya}. 
Hence, this observable has the potential to probe and constrain New Physics contributions to $\Delta a_e$ of order $10^{-13}$. 
 Nevertheless, our analysis has shown that the present (2,2) ISS framework leads to contributions 
to the electron anomalous
magnetic moment of the order of
$|\Delta{a}_e|\sim 10^{-17}$,  which are too small compared to the present value.

\begin{figure}[t]
\begin{center}
\includegraphics[scale=0.6]{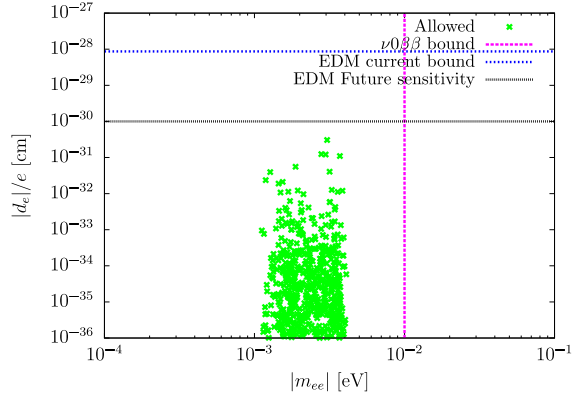}
\caption{Electron EDM vs the effective neutrino mass $|m_{ee}|$. All the points displayed comply with the  constraints discussed in Section~\ref{sec:constraint}. The violet line denotes  the
 conservative limit $|m_{ee}|\leq10^{-2}~\mathrm{eV}$ for $0\nu\beta\beta$
 decay.}
\label{fig:result2}
\end{center}
\end{figure}

Finally, we have considered the contribution of the present model to the effective mass $m_{ee}$
(for the neutrinoless double beta decay). In the present scenario (characterised by a normal hierarchy of the light neutrino spectrum) we have found that the effective neutrino mass lies in  the range $1\lesssim|m_{ee}|/\left(10^{-3}~\mathrm{eV}\right)\lesssim4$, below the conservative experimental limit
$|m_{ee}|\leq10^{-2}~\mathrm{eV}$( see Section \ref{sec:constraint}), 
  as can be seen in Fig.~\ref{fig:result2}.

\section{Discussion}\label{Sec:
Conclusions}

\subsection{Adding more sterile fermions}
In the above analysis, we considered the  Inverse Seesaw  mechanism for neutrino mass generation with the minimal number of RH and sterile states. 
If more sterile fermions are added, one could expect an enhancement of the electron EDM induced by the extra neutrinos in the loop. In order to have an estimation of the possible increase we consider the case in which  the SM is extended by $N$ right-handed neutrinos and $N$ sterile fermions\footnote{As shown in \cite{Abada:2014vea}, the next-to-minimal ISS realisation involving  2 RH neutrinos and 3 sterile fermions leads to the addition of a mass state at the LNV  scale, $\mathcal{O}(\mu)$, to the spectrum obtained in the (2,2) ISS. The latter state is too light to provide an enhancement to the charged lepton EDMs. This statement can be generalised for a number of sterile states larger that the number of RH neutrinos: one finds a spectrum composed by pseudo-Dirac pairs and lighter (mostly) sterile states with  masses $\mathcal{O}(\mu)$.}. 
 Assuming that all the mixing matrix elements $U_{\alpha i}$ are of the same order, i.e. $U_{\alpha i} \simeq 
 \mathcal{O}(U_\mathrm{av})$, and that the loop function $I_D^\prime$ can be approximated by a constant (cf. left panel of Fig.~\ref{fig:j_fac}, in the regime of a strong hierarchy between the pseudo-Dirac pairs), 
one can rewrite the EDM expression  of  Eq.~(\ref{eq:edm2}) for an $(N,N)$ ISS  
realisation  as
\begin{equation}
|d_e^{(N,N)}|\sim
\frac{g_2^4\,e\,m_e}{2(4\pi)^2m_W^2}\Bigl|4N(N-1)\mathrm{Im}
\left(U_\mathrm{av}^4\right)I_D^\prime\Bigr|.
\label{eq:edm_n}
\end{equation}
On the other hand, the new contributions are subject to the constraints discussed on Section~\ref{sec:constraint}, among them $\mu\to e\gamma$. Under the above assumptions,  the branching ratio for  $\mu\to{e}\gamma$ increases,  being approximately given by 
\begin{equation}
\mathrm{Br}\left(\mu\to{e}\gamma\right)\sim
\frac{\sqrt{2}\ G_F^2\ m_\mu^5}{\Gamma_\mu}\left(2N\right)^2
\left|U_\mathrm{av}\right|^4\leq5.7\times10^{-13}.
\label{eq:meg_n}
\end{equation}
Combining Eqs.~(\ref{eq:edm_n}) and (\ref{eq:meg_n}), the current experimental limit on the above branching ratio leads to 
an  upper bound for the electron EDM obtained in or the $(N,N)$ inverse seesaw realisation given by
\begin{equation}
|d_e^{(N,N)}|\lesssim
\left(5.7\times10^{-13}\right)
\frac{e\,m_e\,m_W^2\,\Gamma_\mu}{4\sqrt{2}\pi^2m_\mu^5}
\left(1-\frac{1}{N}\right)\sin\theta |I_D^\prime|,
\end{equation}
where $\sin\theta$ is the phase defined by
$\mathrm{Im}\left(U_\mathrm{av}^4\right)\equiv|U_\mathrm{av}|^4\sin\theta$.

When compared with the (2,2) ISS case, the ($N,N$) ISS contribution to the electron EDM is given by 
\begin{equation}
\left|\frac{d_e^{(N,N)}}{d_e^{(2,2)}}\right|\lesssim
2\left(1-\frac{1}{N}\right).
\end{equation}
For example, if we consider the $N=3$ case, a factor 
 $4/3$ enhancement is expected compared to the results found for the (2,2) ISS realisation. 
In the limit of $N\gg 1$, one obtains, at most,  a factor 2 enhancement, which would nevertheless lead to contributions $\sim10^{-30}~\mathrm{cm}$, thus within 
the optimistic sensitivity of the future experiment by ACME Collaboration.
\subsection{Resonant leptogenesis}
The baryon asymmetry of the Universe is another 
observation pointing towards New Physics scenarios.
 The current center value of the baryon asymmetry, as determined by the PLANCK
Collaboration is given by~\cite{Ade:2013zuv}
\begin{equation}
\frac{n_b-n_{\overline{b}}}{s}=8.59\times10^{-11},
\end{equation}
where $n_b-n_{\overline{b}}$ is the asymmetry between the number density
of baryons and anti-baryons, and $s$ is the entropy density. 
In inverse seesaw realisations, it might be possible to generate the BAU 
 via thermal leptogenesis. A viable leptogenesis is possible in the presence of out-of equilibrium processes violating CP and lepton number. 
 In the following, we  discuss whether or not the regimes leading to a large electron EDM can  also offer the interesting possibility of generating the BAU via leptogenesis. 

Despite being a  low-scale seesaw realisation, a successful baryogenesis could be achieved in the ISS via resonant leptogenesis~\cite{Pilaftsis:2003gt} 
 since the spectrum contains pseudo-Dirac pairs of nearly degenerate states. 
Even if lepton number violation and CP violation are present - in particular a significant  amount of CP violation potentially leading to a large electron EDM -, it turns out that it is not possible to have a successful leptogenesis as one cannot satisfy the out-of-equilibrium condition. 
The decay width of the lightest heavy sterile neutrino $\nu_4$ (whose decays are assumed to be responsible for the generation of a lepton asymmetry) obtained for its dominant channel $\nu_4\to\ell_\alpha^\pm W^\mp$ is given by
\begin{equation}\label{decay-width}
\Gamma_{\nu_4}=\frac{g_2^2\,m_4^3}{16\pi m_W^2}\sum_{\alpha}|U_{\alpha 4}|^2.
\end{equation}
The out-of-equilibrium condition corresponds to having a decay width smaller than the expansion rate of the Universe, which translates into the following inequality,
\begin{equation}\label{condition}
\Gamma_{\nu_4}<\left.H(T)\right|_{T=m_4},
\end{equation}
where $H$ is the Hubble parameter given by
$H\approx1.66\sqrt{g_*}T^2/m_\mathrm{Pl}$ for the radiation-dominated
epoch,  with $g_*$ the effective degrees of freedom of relativistic particles,  $T$ the temperature of the universe and the Planck mass given by 
$m_\mathrm{Pl}=1.22\times10^{19}~\mathrm{GeV}$. 
From Eqs. (\ref{decay-width}) and (\ref{condition}), one can obtain the required order of magnitude for
the mixing angle $|U_{\alpha 4}|$,
\begin{equation}
\sum_\alpha|U_{\alpha 4}|^2\sim10^{-15}
\left(\frac{1~\mathrm{TeV}}{m_4}\right)
\left(\frac{g_*}{100}\right)^{1/2}.
\end{equation}
Thus very small mixing angles are needed to satisfy the out-of-equilibrium condition.  
This means that the  amount of baryon asymmetry generated by resonant leptogenesis
becomes quite small if  large mixings (maximally
$|U_{\alpha i}|\sim10^{-3}$) are  assumed, as required  to induce a large electron EDM (see Section~\ref{Sec:Num}).
If additional sterile states are added, it might be possible to have a successful BAU consistent with a large electron EDM since 
one of the pseudo-Dirac pairs 
of the sterile states is responsible for
generating the BAU,  while  the other pairs can account for large contributions to  the electron EDM.

\section{Summary}\label{Sec: Summary}

We have considered the contribution of sterile neutrinos to the electric dipole
 moment of charged leptons in the (2,2) realisation of the
Inverse Seesaw mechanism.  We have shown that the two pairs of (heavy) pseudo-Dirac mass eigenstates can
 give significant contributions to the electron EDM, close to the future 
experimental sensitivity of ACME,   if their masses are above the electroweak scale. We have further investigated whether or not 
 a successful leptogenesis can be accommodated in this framework,  compatible with a large
 electron EDM. Although such a possibility is precluded for the minimal (2,2) ISS realisation, we do not dismiss its viability in an ISS framework with an extended spectrum ($N,N$) ISS, where $N> 2$.


\section*{Acknowledgments}
We acknowledge partial support from the European Union Horizon 2020
research and innovation programme under the Marie Sk{\l}odowska-Curie: RISE
InvisiblesPlus (grant agreement No 690575)  and 
the ITN Elusives (grant agreement No 674896).
T. T. acknowledges support from P2IO Excellence Laboratory (LABEX).


\section*{Appendix: Loop Functions}

The analytic expression of the loop function $I_D(x_i,x_j)$ has been
derived by FeynCalc in the limit of $x_{\alpha,\beta}\ll1$
and $x_{i,j}\gg1$. 
The loop function represents the contributions of  the two diagrams displayed in Fig.~\ref{fig:edm}. 
Furthermore, each contribution of the diagrams of Fig.~\ref{fig:edm} is 
decomposed into two integral pieces. As a result, the loop function
$I_D(x_i,x_j)$ can be written as 
\begin{equation}
I_D(x_i,x_j)=
I_{D1}^L(x_i,x_j)+I_{D2}^L(x_i,x_j)
+I_{D1}^R(x_i,x_j)+I_{D2}^R(x_i,x_j).
\label{eq:loop_app}
\end{equation}
The superscripts $L,R$ correspond to  the contribution of the left and
right diagrams of Fig.~\ref{fig:edm}, while  the indices $1$ and $2$  refer to two different types of integrals as detailed below. 
Finally it is convenient to anti-symmetrize the loop function (as discussed in Section~\ref{Sec:EDM}) in terms
of $x_i$ and $x_j$ as 
\begin{equation}
I_D(x_i,x_j)\to
\frac{1}{2}\left(
I_D(x_i,x_j)-I_D(x_j,x_i)
\right),
\end{equation}
since only the anti-symmetric part of the loop
function contributes to the charged lepton EDM. 
The terms in the right hand side in Eq.~(\ref{eq:loop_app}) are given by
the integral in terms of the Feynman parameters $s_A$, $t_B$ as 
\begin{equation}
I_{Dn}^{L/R}(x_i,x_j)
=
\int_0^1\prod_{A=1}^{5}ds_A\delta\left(\sum_{A=1}^5s_A-1\right)
\int_0^1\prod_{B=1}^{3+\delta_{2n}}dt_B\delta\left(\sum_{B=1}^{3+\delta_{2n}}t_B-1\right)
F_n^{L/R}(x_i,x_j),
\end{equation}
where the integrands $F_n^{L/R}(x_i,x_j)$ are given by
\begin{equation}
F_n^{L/R}(x_i,x_j)=\frac{N_n^{L/R}(x_i,x_j)}{D^{L/R}(x_i,x_j)}.
\end{equation}
The denominators $D^L(x_i,x_j)$, $D^R(x_i,x_j)$ are given by 
\begin{eqnarray}
D^L(x_i,x_j)
\hspace{-0.2cm}&=&\hspace{-0.2cm}
t_1(s_4+s_5)(s_4+s_5-1)-(1-t_1)(s_1+s_2x_i+s_3x_j),\\
D^R(x_i,x_j)
\hspace{-0.2cm}&=&\hspace{-0.2cm}
-(1-t_1)(s_1+s_4+s_5+s_2x_i+s_3x_j),
\end{eqnarray}
and the numerators $N_n^L(x_i,x_j)$, $N_n^R(x_i,x_j)$ are given by
\begin{eqnarray}
N_{1}^L(x_i,x_j)\hspace{-0.2cm}&=&\hspace{-0.2cm}
\frac{2(1-t_1)^2(s_2-s_3)s_1}{(s_4+s_5)(s_4+s_5-1)}
+\frac{2(-s_2+s_4+s_5-1)}{(s_4+s_5)(s_4+s_5-1)^2}x_i\nonumber\\
&&\hspace{-0.2cm}
+\frac{(1-t_1)\Bigl(3(s_4+s_5)^2+(6s_3-7)(s_4+s_5)-2s_2-6s_3+4\Bigr)}
{(s_4+s_5)(s_4+s_5-1)^2}x_i\nonumber\\
&&\hspace{-0.2cm}
+
\frac{(s_2-s_3)s_1(1-t_1)^2}{2(s_4+s_5)(s_4+s_5-1)^3}x_ix_j
+
\frac{(s_2-s_3)\Bigl(-3(s_4+s_5)t_1+2t_1+1\Bigr)}{2(s_4+s_5)(s_4+s_5-1)^2}x_ix_j,
\\
N_{2}^L(x_i,x_j)\hspace{-0.2cm}&=&\hspace{-0.2cm}
-\frac{4(s_2-s_3)s_1t_1^2}{(s_4+s_5)(s_4+s_5-1)}x_ix_j
-
\frac{4(s_2-s_3)s_1}{(s_4+s_5)^2(s_4+s_5-1)^2}\nonumber\\
&&\hspace{-0.2cm}
-
\frac{(s_2-s_3)s_1}{(s_4+s_5)^2(s_4+s_5-1)^2}x_ix_j
-
\frac{(s_2-s_3)\Bigl(-3(s_4+s_5)t_1+1\Bigr)}{(s_4+s_5)^2(s_4+s_5-1)}x_ix_j
\nonumber\\
&&\hspace{-0.2cm}
+
\frac{-2\Bigl(-3t_1(s_4+s_5)^2+(s_4+s_5)(-6t_1s_3+3t_1+2)-2s_2\Bigr)}
{(s_4+s_5)^2(s_4+s_5-1)}x_i,
\end{eqnarray}
and
\begin{eqnarray}
N_{1}^R(x_i,x_j)\hspace{-0.2cm}&=&\hspace{-0.2cm}
\frac{2s_1(s_2-s_3)t_1(1-t_1)}{(s_4+s_5)(s_4+s_5-1)^2}
+
\frac{2s_1(1-t_1)}{(s_4+s_5)(s_4+s_5-1)^2}x_i\nonumber\\
&&+
\frac{6s_1(s_2-s_3)(1-t_1)\Bigl(1-(s_4+s_5)t_1\Bigr)}{(s_4+s_5)(s_4+s_5-1)^3}-
\frac{(1+s_1t_1)(s_2-s_3)(1-t_1)}{2(s_4+s_5)(s_4+s_5-1)^2}x_ix_j,\\
N_{2}^R(x_i,x_j)\hspace{-0.2cm}&=&\hspace{-0.2cm}
\frac{(3-2s_1t_1)(s_2-s_3)(1-t_1)}{(s_4+s_5)(s_4+s_5-1)^2}x_ix_j\nonumber\\
&&+
\frac{2s_1(1-t_1)\Bigl(2t_1(s_4+s_5)(s_2-s_3)+s_4+s_5-2(s_2-s_3)-1\Bigr)}
{(s_4+s_5)(s_4+s_5-1)^3}x_i\nonumber\\
&&-
\frac{4s_1(s_2-s_3)\Bigl(t_1(s_4+s_5)^2(1+4t_1)-6t_1(s_4+s_5)+1\Bigr)}
{(s_4+s_5)^2(s_4+s_5-1)^3}\nonumber\\
&&-
\frac{10s_1(s_2-s_3)\Bigl(-t_1(s_4+s_5)+1\Bigr)\Bigl((s_4+s_5)(3-4t_1)+2\Bigr)}
{(s_4+s_5)^2(s_4+s_5-1)^3}\nonumber\\
&&+
\frac{6(s_4+s_5)(1-t_1)(4s_2+2s_3+3s_4+3s_5-3)}
{(s_4+s_5)^2(s_4+s_5-1)^2}x_i\nonumber\\
&&+
\frac{6\Bigl(-(s_4+s_5)(3s_2+s_3+2s_4+2s_5-1)+2s_2\Bigr)}
{(s_4+s_5)^2(s_4+s_5-1)^2}x_i\nonumber\\
&&+
\frac{-3(s_4+s_5)(s_4+s_5+2s_2-1)(1-t_1)+(s_4+s_5-1)(4s_2-2s_3+s_4+s_5)}
{(s_4+s_5)(s_4+s_5-1)^3}x_i\nonumber\\
&&+
\frac{(s_2-s_3)}{(s_4+s_5)^2(s_4+s_5-1)}x_ix_j\nonumber\\
&&+
\frac{-s_1(s_2-s_3)\Bigl(t_1(s_4+s_5)^2(1+4t_1)-6t_1(s_4+s_5)+1\Bigr)}
{(s_4+s_5)^2(s_4+s_5-1)^4}x_ix_j.
\end{eqnarray}


\begin{thebibliography}{200}

\bibitem{Minkowski:1977sc}
  P.~Minkowski,
  Phys.\ Lett.\ B {\bf 67} (1977) 421.
  
\bibitem{Yanagida:1979as}
  T.~Yanagida,
  Conf.\ Proc.\ C {\bf 7902131} (1979) 95
   [Conf.\ Proc.\ C {\bf 7902131} (1979) 95].
  
\bibitem{GellMann:1980vs} 
  M.~Gell-Mann, P.~Ramond and R.~Slansky,
  Conf.\ Proc.\ C {\bf 790927} (1979) 315 
  [arXiv:1306.4669 [hep-th]].

\bibitem{Glashow:1979nm}
  S.~L.~Glashow,
  NATO Sci.\ Ser.\ B {\bf 61} (1980) 687.
  
\bibitem{Mohapatra:1979ia}
  R.~N.~Mohapatra and G.~Senjanovic,
  Phys.\ Rev.\ Lett.\  {\bf 44} (1980) 912.

\bibitem{Schechter:1980gr}
  J.~Schechter and J.~W.~F.~Valle,
  Phys.\ Rev.\ D {\bf 22} (1980) 2227.

\bibitem{Schechter:1981cv}
  J.~Schechter and J.~W.~F.~Valle,
  Phys.\ Rev.\ D {\bf 25} (1982) 774.


\bibitem{Mohapatra:1986bd} 
  R.~N.~Mohapatra and J.~W.~F.~Valle,
  Phys.\ Rev.\ D {\bf 34} 1642 (1986).

\bibitem{GonzalezGarcia:1988rw}
  M.~C.~Gonzalez-Garcia and J.~W.~F.~Valle,
  Phys.\ Lett.\ B {\bf 216} (1989) 360.

\bibitem{Deppisch:2004fa}
  F.~Deppisch and J.~W.~F.~Valle,
  Phys.\ Rev.\ D {\bf 72} (2005) 036001
  [hep-ph/0406040].
  
\bibitem{Asaka:2005an} 
  T.~Asaka, S.~Blanchet and M.~Shaposhnikov,
  Phys.\ Lett.\ B {\bf 631} (2005) 151 
  [hep-ph/0503065].


\bibitem{Gavela:2009cd}
  M.~B.~Gavela, T.~Hambye, D.~Hernandez and P.~Hernandez,
  JHEP {\bf 0909} (2009) 038
  [arXiv:0906.1461 [hep-ph]].
  
\bibitem{Ibarra:2010xw}
  A.~Ibarra, E.~Molinaro and S.~T.~Petcov,
  JHEP {\bf 1009} (2010) 108
  [arXiv:1007.2378 [hep-ph]].

\bibitem{Abada:2014vea}
  A.~Abada and M.~Lucente,
  Nucl.\ Phys.\ B {\bf 885} (2014) 651
  [arXiv:1401.1507 [hep-ph]].
  
\bibitem{Barr:2003nn}
  S.~M.~Barr,
  Phys.\ Rev.\ Lett.\  {\bf 92} (2004) 101601
  [hep-ph/0309152].

\bibitem{Malinsky:2005bi} 
  M.~Malinsky, J.~C.~Romao and J.~W.~F.~Valle,
  Phys.\ Rev.\ Lett.\  {\bf 95} (2005) 161801 
  [hep-ph/0506296].

\bibitem{Abada:2014zra}
  A.~Abada, G.~Arcadi and M.~Lucente,
  JCAP {\bf 1410} (2014) 001
  [arXiv:1406.6556 [hep-ph]].

\bibitem{Akhmedov:1998qx} 
  E.~K.~Akhmedov, V.~A.~Rubakov and A.~Y.~Smirnov,
  Phys.\ Rev.\ Lett.\  {\bf 81} (1998) 1359 
  [hep-ph/9803255].

\bibitem{Canetti:2012vf}
  L.~Canetti, M.~Drewes and M.~Shaposhnikov,
  Phys.\ Rev.\ Lett.\  {\bf 110} (2013) 6,  061801
  [arXiv:1204.3902 [hep-ph]].
  
\bibitem{Canetti:2012kh}
  L.~Canetti, M.~Drewes, T.~Frossard and M.~Shaposhnikov,
  Phys.\ Rev.\ D {\bf 87} (2013) 093006
  [arXiv:1208.4607 [hep-ph]].

\bibitem{Abada:2015rta} 
  A.~Abada, G.~Arcadi, V.~Domcke and M.~Lucente,
  JCAP {\bf 1511}, no. 11, 041 (2015)
  [arXiv:1507.06215 [hep-ph]].

\bibitem{Canetti:2014dka}
  L.~Canetti, M.~Drewes and B.~Garbrecht,
  Phys.\ Rev.\ D {\bf 90} (2014) 12,  125005
  [arXiv:1404.7114 [hep-ph]].


\bibitem{Hernandez:2015wna}
  P.~Hern\'andez, M.~Kekic, J.~L\'opez-Pav\'on, J.~Racker and N.~Rius,
  JHEP {\bf 1510} (2015) 067
  [arXiv:1508.03676 [hep-ph]].

 
\bibitem{Mueller:2011nm}
  T.~A.~Mueller, D.~Lhuillier, M.~Fallot, A.~Letourneau, S.~Cormon, M.~Fechner, L.~Giot and T.~Lasserre {\it et al.},
  Phys.\ Rev.\ C {\bf 83} (2011) 054615
  [arXiv:1101.2663 [hep-ex]].

\bibitem{Huber:2011wv}
  P.~Huber,
  Phys.\ Rev.\ C {\bf 84} (2011) 024617
   [Erratum-ibid.\ C {\bf 85} (2012) 029901]
  [arXiv:1106.0687 [hep-ph]].

\bibitem{Mention:2011rk}
  G.~Mention, M.~Fechner, T.~Lasserre, T.~A.~Mueller, D.~Lhuillier, M.~Cribier and A.~Letourneau,
  Phys.\ Rev.\ D {\bf 83} (2011) 073006
  [arXiv:1101.2755 [hep-ex]].
  
\bibitem{Aguilar:2001ty}
  A.~A.~Aguilar-Arevalo {\it et al.}  [LSND Collaboration], 
Phys.\ Rev.\ D {\bf 64} (2001) 112007
  [hep-ex/0104049].
 
\bibitem{AguilarArevalo:2007it}
  A.~A.~Aguilar-Arevalo {\it et al.}  [MiniBooNE Collaboration],
  Phys.\ Rev.\ Lett.\  {\bf 98} (2007) 231801
  [arXiv:0704.1500 [hep-ex]].

\bibitem{AguilarArevalo:2010wv}
  A.~A.~Aguilar-Arevalo {\it et al.}  [MiniBooNE Collaboration],
  Phys.\ Rev.\ Lett.\  {\bf 105} (2010) 181801
  [arXiv:1007.1150 [hep-ex]].

\bibitem{Aguilar-Arevalo:2013pmq}
  A.~A.~Aguilar-Arevalo {\it et al.}  [MiniBooNE Collaboration],
  Phys.\ Rev.\ Lett.\  {\bf 110} (2013) 161801
  [arXiv:1207.4809 [hep-ex], arXiv:1303.2588 [hep-ex]].
  
\bibitem{Acero:2007su}
  M.~A.~Acero, C.~Giunti and M.~Laveder,
  Phys.\ Rev.\ D {\bf 78} (2008) 073009
  [arXiv:0711.4222 [hep-ph]].

\bibitem{Giunti:2010zu}
  C.~Giunti and M.~Laveder,
  Phys.\ Rev.\ C {\bf 83} (2011) 065504
  [arXiv:1006.3244 [hep-ph]].
  
\bibitem{Kopp:2013vaa}
  J.~Kopp, P.~A.~N.~Machado, M.~Maltoni and T.~Schwetz,
  JHEP {\bf 1305} (2013) 050
  [arXiv:1303.3011 [hep-ph]].

\bibitem{Arganda:2014dta}
  E.~Arganda, M.~J.~Herrero, X.~Marcano and C.~Weiland,
  Phys.\ Rev.\ D {\bf 91} (2015) no.1,  015001
  [arXiv:1405.4300 [hep-ph]].

  
\bibitem{Deppisch:2015qwa}
  F.~F.~Deppisch, P.~S.~Bhupal Dev and A.~Pilaftsis,
  New J.\ Phys.\  {\bf 17} (2015) no.7,  075019
  [arXiv:1502.06541 [hep-ph]].

\bibitem{Banerjee:2015gca}
  S.~Banerjee, P.~S.~B.~Dev, A.~Ibarra, T.~Mandal and M.~Mitra,
  Phys.\ Rev.\ D {\bf 92} (2015) 075002
  [arXiv:1503.05491 [hep-ph]].
  
\bibitem{Illana:2000ic}
  J.~I.~Illana and T.~Riemann,
  Phys.\ Rev.\ D {\bf 63} (2001) 053004
  [hep-ph/0010193].

\bibitem{Abada:2014cca}
  A.~Abada, V.~De Romeri, S.~Monteil, J.~Orloff and A.~M.~Teixeira,
  JHEP {\bf 1504} (2015) 051
  [arXiv:1412.6322 [hep-ph]].

\bibitem{Abada:2015zea}
  A.~Abada, D.~Be\v{c}irevi\'c, M.~Lucente and O.~Sumensari,
  Phys.\ Rev.\ D {\bf 91} (2015) no.11,  113013
  [arXiv:1503.04159 [hep-ph]].



\bibitem{Shrock:1980vy}
  R.~E.~Shrock,
  Phys.\ Lett.\ B {\bf 96} (1980) 159.
  
  
\bibitem{Shrock:1980ct}
  R.~E.~Shrock,
  Phys.\ Rev.\ D {\bf 24} (1981) 1232.



\bibitem{Atre:2009rg}
  A.~Atre, T.~Han, S.~Pascoli and B.~Zhang,
  JHEP {\bf 0905} (2009) 030
  [arXiv:0901.3589 [hep-ph]].
  
  
\bibitem{Abada:2012mc}
  A.~Abada, D.~Das, A.~M.~Teixeira, A.~Vicente and C.~Weiland,
  JHEP {\bf 1302} (2013) 048
  [arXiv:1211.3052 [hep-ph]].
  
  \bibitem{Abada:2013aba}
  A.~Abada, A.~M.~Teixeira, A.~Vicente and C.~Weiland,
  JHEP {\bf 1402} (2014) 091
  [arXiv:1311.2830 [hep-ph]].

  
\bibitem{deGouvea:2005jj}
  A.~de Gouvea and S.~Gopalakrishna,
  Phys.\ Rev.\ D {\bf 72} (2005) 093008
  [hep-ph/0508148].
  
\bibitem{Abada:2014nwa}
  A.~Abada, V.~De Romeri and A.~M.~Teixeira,
  JHEP {\bf 1409} (2014) 074
  [arXiv:1406.6978 [hep-ph]].
  
\bibitem{Abada:2015trh}
  A.~Abada and T.~Toma,
  JHEP {\bf 1602} (2016) 174
  [arXiv:1511.03265 [hep-ph]].



\bibitem{Ma:2009gu}
  E.~Ma,
  Phys.\ Rev.\ D {\bf 80} (2009) 013013
  [arXiv:0904.4450 [hep-ph]].

  
\bibitem{Gronau:1984ct}
  M.~Gronau, C.~N.~Leung and J.~L.~Rosner,
  Phys.\ Rev.\ D {\bf 29} (1984) 2539.

\bibitem{Gonzalez-Garcia:2014bfa} 
  M.~C.~Gonzalez-Garcia, M.~Maltoni and T.~Schwetz,
  JHEP {\bf 1411} (2014) 052 
  [arXiv:1409.5439 [hep-ph]].


\bibitem{Antusch:2008tz}
  S.~Antusch, J.~P.~Baumann and E.~Fernandez-Martinez,
  Nucl.\ Phys.\ B {\bf 810} (2009) 369  [arXiv:0807.1003 [hep-ph]].
  
\bibitem{Antusch:2014woa}
  S.~Antusch and O.~Fischer,
  JHEP {\bf 1410} (2014) 094
  [arXiv:1407.6607 [hep-ph]].
   
\bibitem{Benes:2005hn}
  P.~Benes, A.~Faessler, F.~Simkovic and S.~Kovalenko,
  Phys.\ Rev.\ D {\bf 71} (2005) 077901
  [hep-ph/0501295].
 
\bibitem{Blennow:2010th}
  M.~Blennow, E.~Fernandez-Martinez, J.~Lopez-Pavon and J.~Menendez,
  JHEP {\bf 1007} (2010) 096
  [arXiv:1005.3240 [hep-ph]].
  
   \bibitem{Agostini:2013mzu}
M.~Agostini {\it et al.}  [GERDA Collaboration],
Phys.\ Rev.\ Lett.\  {\bf 111} (2013) 12,  122503
[arXiv:1307.4720 [nucl-ex]].

\bibitem{Auger:2012ar} 
  M.~Auger {\it et al.}  [EXO Collaboration],
  Phys.\ Rev.\ Lett.\  {\bf 109} (2012) 032505 
  [arXiv:1205.5608 [hep-ex]].
 
\bibitem{Albert:2014awa} 
  J.~B.~Albert {\it et al.}  [EXO-200 Collaboration],
  Nature {\bf 510} (2014) 229-234 
  [arXiv:1402.6956 [nucl-ex]].
  
  \bibitem{Gando:2012zm} 
  A.~Gando {\it et al.}  [KamLAND-Zen Collaboration],
  Phys.\ Rev.\ Lett.\  {\bf 110} (2013) 062502 
  [arXiv:1211.3863 [hep-ex]].

 
\bibitem{Chanowitz:1978mv}
  M.~S.~Chanowitz, M.~A.~Furman and I.~Hinchliffe,
  Nucl.\ Phys.\ B {\bf 153} (1979) 402.
  
\bibitem{Durand:1989zs}
  L.~Durand, J.~M.~Johnson and J.~L.~Lopez,
  Phys.\ Rev.\ Lett.\  {\bf 64} (1990) 1215.
  
\bibitem{Korner:1992an}
  J.~G.~Korner, A.~Pilaftsis and K.~Schilcher,
  Phys.\ Lett.\ B {\bf 300} (1993) 381
  [hep-ph/9301290].
  
\bibitem{Bernabeu:1993up}
  J.~Bernabeu, J.~G.~Korner, A.~Pilaftsis and K.~Schilcher,
  Phys.\ Rev.\ Lett.\  {\bf 71} (1993) 2695
  [hep-ph/9307295].
  
\bibitem{Fajfer:1998px}
  S.~Fajfer and A.~Ilakovac,
  Phys.\ Rev.\ D {\bf 57} (1998) 4219.
  
\bibitem{Ilakovac:1999md}
  A.~Ilakovac,
  Phys.\ Rev.\ D {\bf 62} (2000) 036010
  [hep-ph/9910213].

\bibitem{Abreu:1996pa}
  P.~Abreu {\it et al.} [DELPHI Collaboration],
  Z.\ Phys.\ C {\bf 74} (1997) 57
   [Erratum: Z.\ Phys.\ C {\bf 75} (1997) 580].
\bibitem{Adriani:1992pq}
  O.~Adriani {\it et al.} [L3 Collaboration],
  Phys.\ Lett.\ B {\bf 295} (1992) 371.

\bibitem{Asaka:2014kia}
  T.~Asaka, S.~Eijima and K.~Takeda,
  Phys.\ Lett.\ B {\bf 742} (2015) 303
  [arXiv:1410.0432 [hep-ph]].


\bibitem{Adam:2013mnn}
  J.~Adam {\it et al.} [MEG Collaboration],
  Phys.\ Rev.\ Lett.\  {\bf 110} (2013) 201801
  [arXiv:1303.0754 [hep-ex]].

\bibitem{Bellgardt:1987du}
  U.~Bellgardt {\it et al.} [SINDRUM Collaboration],
  Nucl.\ Phys.\ B {\bf 299} (1988) 1.

\bibitem{Bertl:2006up}
  W.~H.~Bertl {\it et al.} [SINDRUM II Collaboration],
  Eur.\ Phys.\ J.\ C {\bf 47} (2006) 337.
  
\bibitem{Dev:2013wba}
  P.~S.~B.~Dev, A.~Pilaftsis and U.~k.~Yang,
  Phys.\ Rev.\ Lett.\  {\bf 112} (2014) no.8,  081801
  [arXiv:1308.2209 [hep-ph]].
\bibitem{Das:2014jxa}
  A.~Das, P.~S.~Bhupal Dev and N.~Okada,
  Phys.\ Lett.\ B {\bf 735} (2014) 364
  [arXiv:1405.0177 [hep-ph]].
  
\bibitem{Gavela:1981ri}
  M.~B.~Gavela, G.~Girardi, C.~Malleville and P.~Sorba,
  Nucl.\ Phys.\ B {\bf 193} (1981) 257.

\bibitem{Belanger:2003sd}
  G.~Belanger, F.~Boudjema, J.~Fujimoto, T.~Ishikawa, T.~Kaneko, K.~Kato
	and Y.~Shimizu,
  Phys.\ Rept.\  {\bf 430} (2006) 117
  [hep-ph/0308080].




\bibitem{Baron:2013eja} 
  J.~Baron {\it et al.} [ACME Collaboration],
  Science {\bf 343}, 269 (2014)
  [arXiv:1310.7534 [physics.atom-ph]].

\bibitem{acme:next_generation}
  W.~C.~Griffith, 
  Plenary talk at ``Interplay between Particle \& Astroparticle
  physics 2014'', 
  \url{https://indico.ph.qmul.ac.uk/indico/conferenceDisplay.py?confId=1}. 


\bibitem{Mertig:1990an}
  R.~Mertig, M.~Bohm and A.~Denner,
  Comput.\ Phys.\ Commun.\  {\bf 64} (1991) 345.




\bibitem{Agashe:2014kda}
  K.~A.~Olive {\it et al.} [Particle Data Group Collaboration],
  Chin.\ Phys.\ C {\bf 38} (2014) 090001.


\bibitem{Aboubrahim:2014hya}
  A.~Aboubrahim, T.~Ibrahim and P.~Nath,
  Phys.\ Rev.\ D {\bf 89} (2014) no.9,  093016
  [arXiv:1403.6448 [hep-ph]].


\bibitem{Ade:2013zuv}
  P.~A.~R.~Ade {\it et al.} [Planck Collaboration],
  Astron.\ Astrophys.\  {\bf 571} (2014) A16
  [arXiv:1303.5076 [astro-ph.CO]].

\bibitem{Pilaftsis:2003gt}
  A.~Pilaftsis and T.~E.~J.~Underwood,
  Nucl.\ Phys.\ B {\bf 692} (2004) 303
  [hep-ph/0309342].




\end{thebibliography}
\end{document}